\documentclass[journal]{IEEEtran}

\usepackage{xcolor,soul,framed} 
\usepackage{amsfonts} 
\usepackage{multicol}
\usepackage{multirow}
\usepackage{booktabs}
\usepackage{hyperref} 
\colorlet{shadecolor}{yellow}
\usepackage[pdftex]{graphicx}
\graphicspath{{../pdf/}{../jpeg/}}
\DeclareGraphicsExtensions{.pdf,.jpeg,.png}
\usepackage[numbers,sort&compress]{natbib}
\usepackage[cmex10]{amsmath}
\usepackage{array}
\usepackage{mdwmath}
\usepackage{lscape}
\usepackage{mdwtab}
\usepackage{eqparbox}
\usepackage{url}
\usepackage{float}
\usepackage{bm}
\usepackage{lineno}
\usepackage{makecell}
\usepackage{url}
\usepackage{tikz}
\usetikzlibrary{positioning}
\usepackage{calc}
\usepackage{diagbox}
\usepackage{url}
\usepackage{threeparttable}
\usepackage{verbatim}
\usepackage{appendix}
\usepackage[figuresright]{rotating}
\usepackage{amsfonts,amssymb}
\begin{document}
    \title{Learnable Linguistic Watermarks for Tracing Model Extraction Attacks on Large Language Models}
  \author{Minhao Bai, Kaiyi Pang, Yongfeng Huang
    }



\maketitle

\begin{abstract}
In the rapidly evolving domain of artificial intelligence, safeguarding the intellectual property of Large Language Models (LLMs) is increasingly crucial. Current watermarking techniques against model extraction attacks, which rely on signal insertion in model logits or post-processing of generated text, remain largely heuristic. We propose a novel method for embedding learnable linguistic watermarks in LLMs, aimed at tracing and preventing model extraction attacks. Our approach subtly modifies the LLM's output distribution by introducing controlled noise into token frequency distributions, embedding an statistically identifiable controllable watermark.We leverage statistical hypothesis testing and information theory, particularly focusing on Kullback-Leibler Divergence, to differentiate between original and modified distributions effectively. Our watermarking method strikes a delicate well balance between robustness and output quality, maintaining low false positive/negative rates and preserving the LLM's original performance.

\end{abstract}


\IEEEpeerreviewmaketitle

\section{Introduction}

It is a feasible method to fine-tune Large Language Models (LLMs) \cite{BERT,GPT2,OPT,LLAMA,ChatGLM,GPT3} by aligning them with human prompts. This process involves expert knowledge and manual annotations, which can be costly. To address these challenges, practitioners sometimes train LLMs on texts generated by pre-instructed models\cite{alpaca,vicuna}. This technique has been used in various works to create instruction data for recent LLMs, and it can also occur unintentionally when using LLM outputs for their tasks. This raises concerns about whether the fine-tuned model is considered a derivative work of the original model. Sometimes the model trainers will open their data sources faithfully, but they can also report disingenuous data sources or even hide it. Hence, there is a need to understand how to identify the use of LLM outputs as training data.


Detecting model-generated texts has become increasingly challenging, particularly in cases where the entropy of generated content is not sufficient. 
The mainstream method to tackle this issue is through watermarking\cite{undetectable,unigram,redgreenlist,expedit,rainbowlist,semanticinvarient,qu2024provably,blackbox1,whitebox1,blackbox2}, which involves embedding a hidden message within the text during or after the generation process to identify the generating model. While watermarking has been used for thousands of years, there is a growing interest in its application to LLMs, since recent techniques enables efficient detection without significantly compromising the quality of the output. 

A watermark method includes the injection and detection procedures. The injection procedure may involve the modification of produced texts or the model itself. Most of the watermark method that only modify the produced texts without using the language model can be seen as black-box watermark\cite{blackbox1,redgreenlist,blackbox2,undetectable}. Some watermarks only require the access to the model's prediction, which can be seen as grey-box watermark\cite{undetectable,unigram,redgreenlist,expedit,rainbowlist,semanticinvarient,qu2024provably}. And the watermarks that get access to the all parameters of model is called white-box watermark\cite{whitebox1,whitebox2}.

Current black-box watermark methods\cite{blackbox1,redgreenlist,blackbox2,undetectable} often face tough detection problems, since the redundant space left for watermark is not quite enough. And the straightforward modification of generated texts is hard to control. White-box watermarks\cite{whitebox1,whitebox2} are not the focus of mainstream since the access to the whole parameters of model sounds too offensive. The training process is not that stable and the changing of parameter is not likely to be accepted.

Therefore, most of watermark research focus on the grey-box scenario\cite{undetectable,unigram,redgreenlist,expedit,rainbowlist,semanticinvarient,qu2024provably}. These works aim at the robustness and maintaining the quality of texts. Red-green list \cite{redgreenlist} is a classical resolution, but it is harmful to the quality of generated texts. Unigram \cite{unigram} is based on Red-green list and gives a formal robustness proof. Exp-edit \cite{expedit} provides a distortion-free watermark that does not explicitly lower the quality. Miranda Christ et al \cite{undetectable} construct an undetectable watermark that watermarked texts could not be distinguished from the original texts. However, in the scenario of tracking model extraction attack, the watermark should be learnable, which is contradict to undetectable.

Our work centres on tracking model extraction attacks while achieving stable and robust watermark detection without degrading the quality of the generated text. 

\section{Method}
\subsection{Sample the Frequency and Noise It}
First, we can choose a dataset $D$ which is a bunch of examples that obey the distribution $\mathcal{P_D}$. Then we accumulate the frequency of each token, as following Equation \ref{fre} shows:
\begin{equation}\label{fre}
    F_D(w_i) = \mathbb{E}_{D\sim\mathcal{P_D}}[\mathcal{P_D}(w_i)]
\end{equation}
where $F_D$ stands for the frequency computed in dataset$D$ and $ w_i$ represents a token. In practice it can be estimated by Equation \ref{fre_cal}.
\begin{equation}\label{fre_cal}
    F_D(w_i) = \frac{Count(w_i)}{\sum\limits_{w_i \in V} Count(w_i)}
\end{equation}
where $V$ denotes the vocabulary of tokens and $|V|$ represents the size of $V$.

It is feasible to use the model-generated text as a dataset or a human corpus, which is not a matter. The next step is to add gaussian noise (or any other form of noise you like) to this frequency.

\begin{equation}
    \Hat{F_D}(w_i) = F_D(w_i) + n_i, n_i \sim \mathcal{N}(0,\sigma^2)
\end{equation}

Now it is time to watermark the output of our protected model. We denote the predict distribution of protected model as $\mathcal{P_{LM}}$. We modify this distribution (denotes as $\mathcal{\hat{P}_{LM}}$) to make it generates a bunch of text whose frequency distribution is $\Hat{F_D}(w_i)$. That requires
\begin{equation}
    \Hat{F}_D(w_i) = \mathbb{E}_{D\sim\mathcal{P_D}}[\mathcal{\hat{P}_{LM}}(w_i)]
\end{equation}
We designed a method to modify the distribution, as Equation \ref{mod} shown:
\begin{equation}\label{mod}
    \mathcal{\hat{P}_{LM}}(w_i) = \mathcal{P_{LM}}(w_i) \times \frac{\hat{F}_D(w_i)}{F_{LM}(w_i)}
\end{equation}

Thus, constructing a modified distribution is done.
The strength of watermark depends on the KL Divergence between $\hat{F}_D$ and $F_{LM}$. Additionally, if protected model's frequency is used to be noised, the strength of watermark depends on the variance of Gaussian noise.

It is necessary to note that the methods of modifying distribution are various. The main purpose of this step is to construct a unique distribution as the detection anchor of our watermark.

\subsection{Distinguish 2 Different Distributions}
After the construction of modified distribution $\mathcal{\hat{P}_{LM}}$, we need to distinguish this distribution generated texts with others.
It is feasible to compute the average information $I$ of chosen tokens, as Equation \ref{info} shown:
\begin{equation}\label{info}
    I_{\mathcal{\hat{P}_{LM}}} = \mathbb{E}[-\log(\mathcal{\hat{P}_{LM}}(w_i))]
\end{equation}
Assuming an arbitrary distribution $\mathcal{P_A}$ that is different from $\mathcal{\hat{P}_{LM}}$, with a KL Divergence $KL(\mathcal{P_A}||\mathcal{\hat{P}_{LM}}) > 0$,  the 
expectation computed under $\mathcal{\hat{P}_{LM}}$ is $I_{\mathcal{{P}_{A}}}=\mathbb{E}_{w_i \sim \mathcal{\hat{P}_{LM}}}[-\log(\mathcal{\hat{P}_{A}}(w_i))]$. There will be a difference between $I_{\mathcal{{P}_{A}}}$ and $I_{\mathcal{\hat{P}_{LM}}}$, as Equation \ref{diff} shown:
\begin{equation}\label{diff}
\begin{split}
   I_{\mathcal{\hat{P}_{LM}}} - I_{\mathcal{{P}_{A}}} & = \mathbb{E}_{w_i \sim \mathcal{\hat{P}_{LM}}}[-\log(\frac{\mathcal{\hat{P}_{LM}}(w_i)}{\mathcal{{P}_{A}}(w_i)})] \\ 
    & = KL(\mathcal{\hat{P}_{LM}}(w_i)||\mathcal{{P}_{A}}(w_i)) > 0
\end{split}
\end{equation}

In most cases the KL Divergence is not big enough for us to precisely distinguish the distributions. So the accumulation of information over the generated tokens is necessary.

With hypothesis testing techniques things will be more clear. We define the hypothesis $\mathcal{H}_0$ and $\mathcal{H}_1$ as follows:
\begin{equation}
    \begin{split}
       &\mathcal{H}_0: Sequence [w_1,w_2,...] \sim \mathcal{\hat{P}_{LM}} \\
       & \mathcal{H}_1: Sequence [w_1,w_2,...] \sim \mathcal{{P}_{A}}
    \end{split}
\end{equation}
The information difference between hypothesis $\mathcal{H}_0$ and $\mathcal{H}_1$ is accumulated over the sequence.
\begin{equation}
    I_{\mathcal{\hat{P}_{LM}}}(\mathcal{H}_0) - I_{\mathcal{{P}_{A}}}(\mathcal{H}_1) \longleftarrow \sum\limits_i -\log(\frac{\mathcal{\hat{P}_{LM}}(w_i)}{\mathcal{{P}_{A}}(w_i)})
\end{equation}
Considering the Type \uppercase\expandafter{\romannumeral1} error rate not greater than $\alpha$, we can define the upper bound $B_I$ to accept hypothesis $\mathcal{H}_0$ as :
\begin{equation}
    Pr([w_1,w_2,...] \sim \mathcal{{P}_{A}}; I_{\mathcal{\hat{P}_{LM}}}(\mathcal{H}_0) - I_{\mathcal{{P}_{A}}}(\mathcal{H}_1) < B_{\uppercase\expandafter{\romannumeral1}}) < \alpha
\end{equation}
Thus the information difference bound $B_I$ and the upper bound of Type \uppercase\expandafter{\romannumeral1} error rate $\alpha$ should satisfy the following equation:
\begin{equation}
    B_{\uppercase\expandafter{\romannumeral1}} \geq -ln(\frac{\alpha}{1-\alpha})
\end{equation} 
Since the information difference is related to the KL Divergence, the expected number of accumulated tokens $N_I$ is
\begin{equation}
    N_{\uppercase\expandafter{\romannumeral1}} \geq -ln(\frac{\alpha}{1-\alpha})/KL(\mathcal{\hat{P}_{LM}}||\mathcal{{P}_{A}})
\end{equation}
That also proves that the KL Divergence stands for the strength of the watermark.

The large KL Divergence means a stronger watermark intensity, but it may degrade the general performance of the origin language model.

Considering the Type \uppercase\expandafter{\romannumeral2} error rate not greater than $\beta$, we can define the lower bound $B_{II}$ to accept hypothesis $\mathcal{H}_1$ as :
\begin{equation}
    Pr([w_1,w_2,...] \sim \mathcal{\hat{P}_{LM}} ; I_{\mathcal{\hat{P}_{LM}}}(\mathcal{H}_0) - I_{\mathcal{{P}_{A}}}(\mathcal{H}_1) > B_{\uppercase\expandafter{\romannumeral2}}) < \beta
\end{equation}
Thus the information difference bound $B_{II}$ and the upper bound of Type \uppercase\expandafter{\romannumeral2} error rate $\beta$ should satisfy the following equation:
\begin{equation}
    B_{II} \geq \ln \frac{1-\beta}{\beta}
\end{equation}

Similarly, the expected number of accumulated tokens $N_{II}$ is
\begin{equation}
    N_{II} \geq \ln \frac{1-\beta}{\beta}/KL(\mathcal{\hat{P}_{LM}}||\mathcal{{P}_{A}})
\end{equation}

Choosing the appropriate decision bound will be a trade-off between expected number of accumulated tokens and false-positive/negative rate (FPR/FNR).

In practice, we will first calculate the accumulated information of the modified distribution generated texts, and then compare this value with that of the texts that will be detected. The computing procedure involves:
\begin{equation}
    I_{\mathcal{\hat{P}_{LM}}} \longleftarrow \frac{1}{T}\sum_{t = 1}^T -\log \hat{P}_{LM}(w_t|w_{<t})
\end{equation}

The sequence $\{w_t\}_{t=1}^T$ that follows the modified distribution will have a lower information than the sequence that not follows it, which is proved before. The decision bound can be set in practice or compute according to the FPR/FNR.

\subsection{Watermark tracing the model extraction}

In the subsections before we talked about the construction of a noised frequency distribution and a theoretical method to distinguish 2 different distributions. This subsection will explain how to construct a learnable watermark. The basic property of watermark includes efficient embedding and detection methods, and in order to tracking the model extraction attack, the watermark should be detectable in the extraction model generated texts.

Therefore, we conclude the property of this type of watermark into 3 pieces: 
\begin{itemize}
    \item Low FPR/FNR. Practical watermarking should keep a reasonable FPR/FNR to ensure the efficiency.
    \item Maintain Quality. It should not cause degradation of model's ability.
    \item Learnable. Models that trained from the watermarked text will generate texts that can be detected as watermarked.
\end{itemize}

As we mentioned before, the problem of FPR/FNR is restricted by the number of tokens. If the access to the extraction model and original model is guaranteed, you can generate texts as long as you like. FPR/FNR is adjustable and easy to achieve. The quality problem may sounds a bit awkward, since the quality is not well-defined. If we talked about perplexity, diversity, or metrics like MAUVE or Bertscore, it is easy to maintain these metrics and inject strong watermark. 

The learnability of watermark will be the vital thing that we focus on. As we modified the distribution of original model, the generated texts will follow the modified distribution. Ideally the model that train on these texts will learn the distribution and generate texts that obey this watermarked distribution. However, the learning procedure of models is not stable and probably has a high level of noise. Since the training models from our dataset is completely controlled by the adversaries, and formally analysing the impact of different training procedures is almost impossible, we will explore the different conditions of training in experiment section. 

At least, we remarked that the KL divergence between the modified distribution $\mathcal{\hat{P}_{LM}}$ and the distribution of extraction model $\mathcal{{P}_{EXT}}$ should be lower than the decision bound. That is the working limits of this type of watermarking.
\bibliographystyle{IEEEtran}
\bibliography{Bibliography,IEEEabrv}
\appendices
\subsection{Explanation of Equation \ref{diff}}
This equation used the Gibbs' inequality. If $\sum_i p_i = 1, \sum_i q_i = 1$, and $ p_i, q_i \in (0,1]$,
\begin{equation}
    \sum_i p_i\log p_i \geq \sum_i p_i \log q_i
\end{equation}
Therefore the KL Divergence will be non-negative, since
\begin{equation}
\begin{split}
    KL(P||Q) & = \sum_i p_i \log \frac{p_i}{q_i}  \\
    & = \sum_i p_i\log p_i - \sum_i p_i \log q_i \geq 0
\end{split}
\end{equation}
If there exists $p_i \neq q_i$, the KL Divergence is strictly greater than 0.
\end{document}